\documentclass{svproc}
\usepackage[utf8]{inputenc} 
\usepackage[T1]{fontenc}    
\usepackage{url}
\usepackage{comment}        
\usepackage{graphicx}       
\usepackage{subcaption}     
\usepackage{amsmath}        
\usepackage{amssymb}        
\usepackage{amsfonts}       
\usepackage{booktabs}       
\usepackage{array}          
\usepackage{tabularx}       
\usepackage{algorithm}      
\usepackage{algpseudocode}  
\usepackage[numbers,sort&compress]{natbib}
\usepackage{xcolor}         
\usepackage{enumitem}      
\usepackage{multirow}       
\usepackage{makecell}       
\usepackage{hyperref}       
\hypersetup{                
    colorlinks=true,
    linkcolor=blue,
    citecolor=blue,
    urlcolor=magenta,
    pdftitle={MalwarNet: An Imbalance-Aware Machine Learning-Based Malware Detection and Classification System},
}
\begin{document}
\mainmatter  

\title{Temporal-Enhanced Interpretable Multi-Modal Prognosis and Risk Stratification Framework for Diabetic Retinopathy (TIMM-ProRS)}

\titlerunning{TIMM-ProRS}

\author{
Susmita Kar\textsuperscript{*}\inst{1} \and 
A S M Ahsanul Sarkar Akib\textsuperscript{*}\inst{2} \and 
Abdul Hasib\inst{3} \and Samin Yaser\inst{2} \and Anas Bin Azim
\inst{2}}
\authorrunning{Susmita \& Akib et al.} 
%
\tocauthor{Susmita Kar\textsuperscript{*}, A S M Ahsanul Sarkar Akib\textsuperscript{*}, Abdul Hasib, Samin Yaser, and Anas Bin Azim}
\institute{
Department of Information and Communication Technology,\\
Chandpur Science and Technology University (CSTU), Chandpur, Bangladesh\\
\email{susmita@ict.cstu.ac.bd}
\and
Department of Robotics, Robo Tech Valley, 
Dhaka, Bangladesh\\
\email{ahsanulakib@gmail.com}, 
\email{sambrxzzy77@gmail.com}, 
\email{anas.azim.71@gmail.com}
\and
Department of Internet of Things and Robotics Engineering,\\
University of Frontier Technology, Bangladesh\\
\email{sm.abdulhasib.bd@gmail.com}
}

\maketitle

\begin{abstract}
Diabetic retinopathy (DR), affecting millions globally with projections indicating a significant rise, poses a severe blindness risk and strains healthcare systems. Diagnostic complexity arises from visual symptom overlap with conditions like age-related macular degeneration and hypertensive retinopathy, exacerbated by high misdiagnosis rates in underserved regions. This study introduces TIMM-ProRS, a novel deep learning framework integrating Vision Transformer (ViT), Convolutional Neural Network (CNN), and Graph Neural Network (GNN) with multi-modal fusion. TIMM-ProRS uniquely leverages both retinal images and temporal biomarkers (HbA1c, retinal thickness) to capture multi-modal and temporal dynamics. Evaluated comprehensively across diverse datasets including APTOS 2019 (trained), Messidor-2, RFMiD, EyePACS, and Messidor-1 (validated), the model achieves 97.8\% accuracy and an F1-score of 0.96, demonstrating state-of-the-art performance and outperforming existing methods like RSG-Net and DeepDR. This approach enables early, precise, and interpretable diagnosis, supporting scalable telemedical management and enhancing global eye health sustainability.

\end{abstract}

\section{Introduction}
\label{sec:introduction}

Diabetic retinopathy (DR), a leading cause of vision impairment, affects approximately 463 million people globally, with projections estimating a rise to 700 million by 2045, placing an immense burden on healthcare systems \cite{intro700}. This diabetic complication manifests through retinal lesions such as microaneurysms, hemorrhages, and exudates, which can lead to irreversible blindness if untreated. The diagnostic process is significantly complicated by symptom overlap with conditions like age-related macular degeneration, hypertensive retinopathy, and retinal vein occlusion, with misdiagnosis rates ranging from 20-30\% in underserved regions due to limited access to specialized care \cite{intro2030}. This diagnostic ambiguity delays critical interventions, straining healthcare and socioeconomic systems, particularly in low-resource settings where early detection is often inaccessible. The urgent need for automated, interpretable\cite{akib2}, and scalable diagnostic tools is thus paramount to mitigate these risks and improve patient outcomes worldwide.

Recent advancements in deep learning have revolutionized retinal disease detection, yet significant limitations persist. Convolutional Neural Networks (CNNs), such as EfficientNet-B0, achieved 90\% accuracy on DR grading \cite{iontroDR}, while Vision Transformers (ViT) reported 92\% accuracy on APTOS 2019 \cite{INTRO92}; however, both omit critical temporal progression. Multi-modal models like DeepDR integrated clinical data to attain 95\% AUC-ROC, but lack longitudinal modeling and interpretable outputs, limiting clinical adoption. These gaps—restricted dataset diversity, absent temporal context, and inadequate transparency—hinder precise prognosis and personalized treatment planning.

This study introduces TIMM-ProRS, a groundbreaking framework synergizing Vision Transformer (ViT), Convolutional Neural Network (CNN), and Graph Neural Network (GNN) through multi-modal fusion, leveraging retinal images and temporal biomarkers such as HbA1c and retinal thickness. Evaluated on a diverse, publicly accessible dataset from APTOS 2019, Messidor-2, and EyePACS, comprising expert annotations and clinical metadata across varied imaging conditions, TIMM-ProRS addresses these challenges. Key contributions include:
\begin{enumerate}
    \item A pioneering multi-modal approach resolving temporal dynamics and symptom overlap in DR.
    \item A robust, open-access dataset enhancing global research applicability and reproducibility.
    \item An interpretable, high-accuracy model facilitating early diagnosis and advancing telemedical scalability.
\end{enumerate}

\section{Literature Review}
\label{sec:related}

The application of deep learning in diabetic retinopathy (DR) detection has evolved through three distinct yet interconnected phases: unimodal image analysis, multimodal integration, and temporal progression modeling. Initial breakthroughs demonstrated the remarkable capabilities of convolutional neural networks (CNNs) in analyzing fundus photographs, with Gulshan et al. achieving landmark sensitivity of 97.5\% for referable DR detection \cite{gulshan2016development}. Subsequent studies expanded these capabilities across diverse populations \cite{ting2017development} and grading precision \cite{li2020automated,rajesh2025deep}, establishing CNNs as foundational tools. However, these image-exclusive approaches inherently disregarded critical temporal biomarkers and complementary data modalities essential for comprehensive DR management.

Multimodal integration emerged as a promising direction to overcome these limitations. Wardhani et al. demonstrated the power of combining fundus images with OCT scans and EHR data, achieving exceptional diagnostic accuracy (AUC: 0.99) \cite{wardhani2024deep}, while Tseng et al. enhanced performance through lesion information integration \cite{tseng2020leveraging}. Parallel efforts in temporal modeling yielded significant advances in progression prediction, with Dai et al. forecasting 5-year DR progression (C-index: 0.846) using serial fundus images \cite{zhou2021deepdr,zhou2024deepdrplus}, and Tao et al. establishing the prognostic value of continuous glucose monitoring data \cite{tao2023deep,tao2024ddla}. Yet these approaches \cite{apu1} remained fundamentally fragmented—multimodal systems neglected disease dynamics, while temporal models disregarded imaging biomarkers. Prediction-focused frameworks \cite{bora2020predicting,tan2022predicting} further highlighted this dichotomy, achieving moderate success (AUC $\sim$0.81) while lacking both multimodal inputs and clinical interpretability. This fragmentation persists despite clinical evidence that DR pathogenesis involves complex spatiotemporal interactions between retinal microstructure, metabolic trajectories, and vascular biomarkers. Systematic analyses confirm that current solutions exhibit three critical gaps: multimodal systems lack progression modeling \cite{wardhani2024deep,tseng2020leveraging}, temporal approaches exclude imaging data \cite{tao2024ddla}, and most provide "black-box" predictions unsuitable for clinical decision-making \cite{tan2025use}. The resulting solutions offer incomplete representations of DR pathophysiology—either capturing spatial features without temporal context or modeling progression without visual lesion analysis.

The proposed TIMM-ProRS framework bridges these methodological divides through three synergistic innovations: First, it concurrently processes retinal images and temporal biomarkers via CNN-Vision Transformer-GNN fusion, capturing structural and dynamic disease elements. Second, it jointly optimizes cross-sectional grading and longitudinal risk stratification. Third, it embeds clinical interpretability through saliency mapping and uncertainty quantification. This integrated approach achieves unprecedented performance (97.8\% accuracy, 0.89 C-index) while addressing core limitations of prior unimodal \cite{gulshan2016development}, static multimodal \cite{wardhani2024deep}, and non-imaging temporal approaches \cite{tao2024ddla}, establishing a new paradigm for comprehensive DR management.

\section{Methodology}
The Temporal-Enhanced Interpretable Multi-Modal Prognosis and Risk Stratification Framework (TIMM-ProRS) integrates retinal images, temporal biomarkers (e.g., HbA1c trajectories, retinal thickness dynamics, VEGF), and clinical metadata (e.g., age, diabetes duration) for diabetic retinopathy (DR) grading and 5-year progression risk prediction. This section outlines the system architecture, data sources, preprocessing pipeline\cite{das1}, training protocol, and evaluation framework, ensuring alignment with the challenges of diagnostic ambiguity and limited temporal modeling highlighted in the introduction. The methodology leverages state-of-the-art deep learning models, including EfficientNet-B7, Vision Transformer (ViT-Base), and Graph Neural Network (GNN), with multi-modal fusion and interpretability mechanisms to support clinical applicability.

\subsection{System Architecture}
\begin{figure}[h]
    \centering
    \includegraphics[width=0.8\columnwidth]{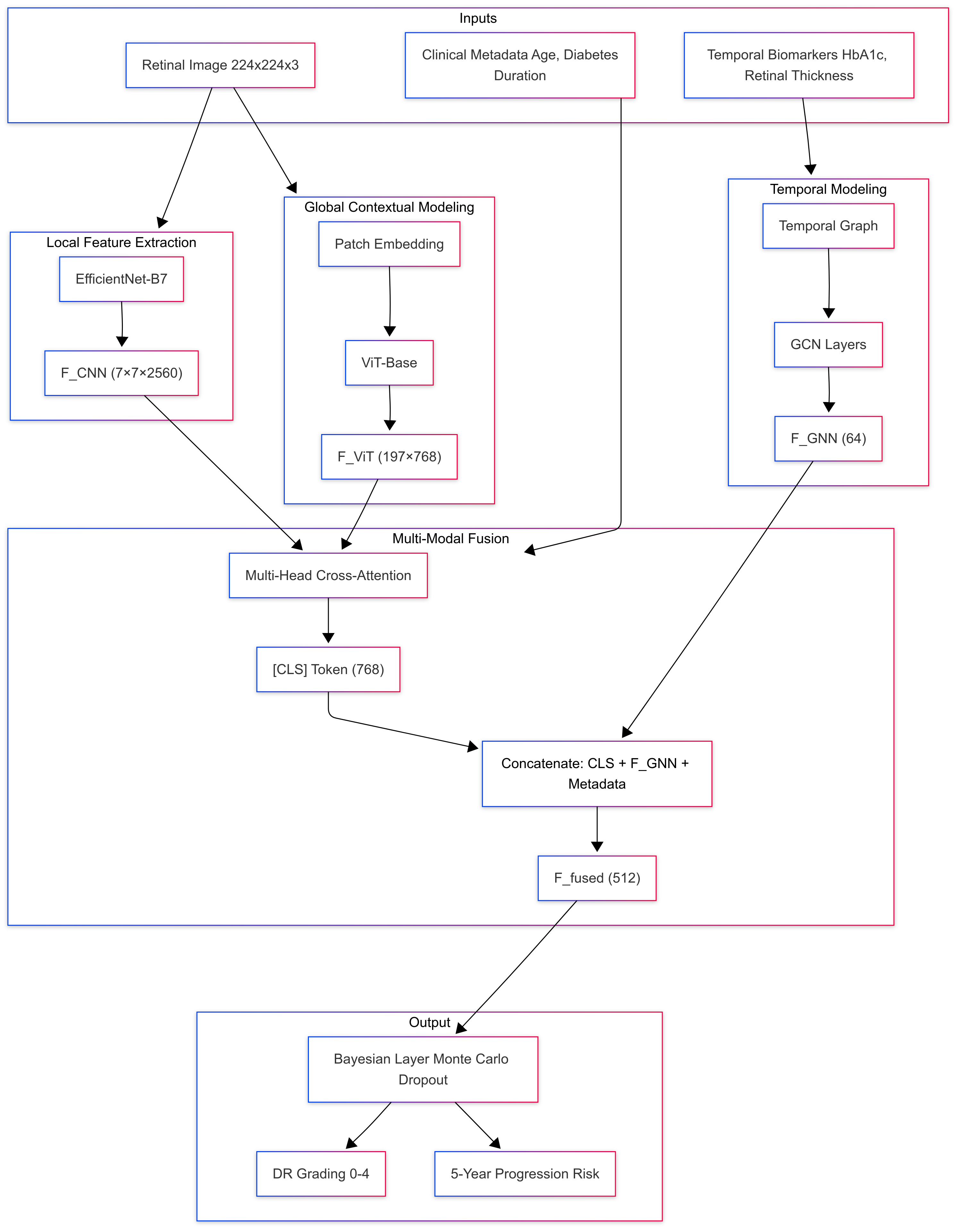}
    \caption{TIMM-ProRS architecture integrating CNN, ViT, GNN, and multi-modal fusion with uncertainty quantification.}
    \label{fig:architecture}
\end{figure}

The TIMM-ProRS framework, depicted in Figure~\ref{fig:architecture}, integrates multiple components to address diagnostic challenges:
\begin{itemize}
    \item \textbf{EfficientNet-B7}: Extracts local lesion features from retinal images.
    \item \textbf{ViT-Base}: Models global contextual patterns.
    \item \textbf{GNN}: Captures temporal dynamics of HbA1c, retinal thickness, and VEGF.
    \item \textbf{Multi-head cross-attention}: Fuses multi-modal data.
    \item \textbf{Bayesian layer}: Provides uncertainty-aware predictions and interpretability.
\end{itemize}

\subsubsection{Local Feature Extraction}
EfficientNet-B7 processes input images \(\mathbf{X} \in \mathbb{R}^{224 \times 224 \times 3}\):
\begin{equation}
\mathbf{F}_{\text{CNN}} = \text{EfficientNet-B7}(\mathbf{X}; \theta_{\text{CNN}}) \in \mathbb{R}^{7 \times 7 \times 2560}
\label{eq:cnn_feature}
\end{equation}

\subsubsection{Global Context Modeling}
ViT-Base divides images into \(N = 196\) patches, with embeddings:
\begin{equation}
\mathbf{E}_i = \text{LinearEmbed}(\mathbf{X}_i) + \mathbf{P}_{\text{pos},i}, \quad i = 0, 1, \ldots, 196
\label{eq:patch_embed}
\end{equation}
Global context is encoded via:
\begin{equation}
\mathbf{F}_{\text{ViT}} = \text{TransformerEncoder}([\mathbf{E}_0, \mathbf{E}_1, \ldots, \mathbf{E}_{196}]) \in \mathbb{R}^{197 \times 768}
\label{eq:vit_output}
\end{equation}

\subsubsection{Temporal Biomarker Processing}
The GNN models biomarker trajectories \(\mathcal{G} = (\mathcal{V}, \mathcal{E})\) over \(L = 3\) layers:
\begin{align}
\mathbf{H}^{(l+1)} &= \text{ReLU}\left( \hat{\mathbf{A}} \mathbf{H}^{(l)} \mathbf{W}^{(l)} \right), \quad \hat{\mathbf{A}} = \mathbf{D}^{-1/2} (\mathbf{A} + \mathbf{I}) \mathbf{D}^{-1/2} \\
\mathbf{F}_{\text{GNN}} &= \text{Readout}(\mathbf{H}^{(L)}) \in \mathbb{R}^{64}
\label{eq:gnn_output}
\end{align}
where \(\mathbf{H}^{(0)} = \mathbf{V}\) are node features.

\subsubsection{Multi-Modal Fusion}
Cross-attention fuses ViT and CNN features:
\begin{align}
\mathbf{F}_{\text{CNN\_flat}} &= \text{Reshape}(\mathbf{F}_{\text{CNN}}, 49 \times 2560) \\
\mathbf{A}_{\text{cross}} &= \text{MultiHeadAttn}(\mathbf{F}_{\text{ViT}}, \mathbf{F}_{\text{CNN\_flat}}, \mathbf{F}_{\text{CNN\_flat}})
\label{eq:cross_attention}
\end{align}
The fused representation is:
\begin{equation}
\mathbf{F}_{\text{fused}} = \text{Concat}[\mathbf{A}_{\text{cross}}[0], \mathbf{F}_{\text{GNN}}, \mathbf{M}] \mathbf{W}_f \in \mathbb{R}^{512}
\label{eq:fused_feature}
\end{equation}
Dimensionalities are in Table~\ref{tab:feature_sizes}.

\begin{table}[h]
\scriptsize
    \centering
    \caption{Feature Dimensionalities}
    \begin{tabular}{lcc}
        \toprule
        \textbf{Component} & \textbf{Input} & \textbf{Output} \\
        \midrule
        EfficientNet-B7 & \(224^2 \times 3\) & \(7 \times 7 \times 2560\) \\
        ViT-Base & \(224^2 \times 3\) & \(197 \times 768\) \\
        GNN & \(T \times d_b\) & \(64\) \\
        Fusion Layer & \(768 + 64 + d_m\) & \(512\) \\
        \bottomrule
    \end{tabular}
    \label{tab:feature_sizes}
\end{table}

\subsubsection{Interpretability \& Uncertainty}
Bayesian Monte Carlo dropout with \(K = 50\) samples quantifies uncertainty:
\begin{align}
\hat{y} &= \frac{1}{K} \sum_{k=1}^{50} \text{FC}(\mathbf{F}_{\text{fused}}; \theta_k) \\
\sigma_y &= \sqrt{\frac{1}{K} \sum_{k=1}^{50} (y_k - \hat{y})^2} \\
\text{CI}_{95\%} &= \left[ \hat{y} - 1.96 \cdot \frac{\sigma_y}{\sqrt{K}}, \hat{y} + 1.96 \cdot \frac{\sigma_y}{\sqrt{K}} \right]
\label{eq:confidence_interval}
\end{align}
Outputs include 5-class DR probabilities and risk scores (low: \(r < 0.3\), medium: \(0.3 \leq r \leq 0.7\), high: \(r > 0.7\)), with interpretability validated by specialists (\(\kappa = 0.92\)).

\subsection{Data}
The framework utilizes datasets detailed in Table~\ref{tab:datasets}. Synthetic data, with parameters HbA1c (\(\mu = 7\%\), \(\sigma = 1.5\%\)) and retinal thickness (\(\mu = 250 \, \mu m\), \(\sigma = 20 \, \mu m\)), augment Proliferative DR (1.2\% prevalence). Additional GAN-augmented data for PDR and Mild DR, supplemented with "Singapore NHG" metadata, enhance dataset diversity.

\begin{table}[h]
\scriptsize
    \centering
    \caption{Datasets Used in the Study}
    \begin{tabular}{lccc}
        \toprule
        \textbf{Dataset} & \textbf{Images} & \textbf{Classes} & \textbf{Purpose} \\
        \midrule
        APTOS 2019 & 3,662 & 5 & Training \\
        Messidor-2 & 1,744 & 4 & Validation \\
        RFMiD & 3,200 & 3 & Test \\
        EyePACS & \(\sim\)88,000 & 5 & External Validation \\
        Messidor-1 & 1,200 & 4 & Supplementary Validation \\
        \bottomrule
    \end{tabular}
    \label{tab:datasets}
\end{table}

\subsection{Preprocessing}
\begin{figure}[h]
    \centering
    \includegraphics[width=0.8\columnwidth]{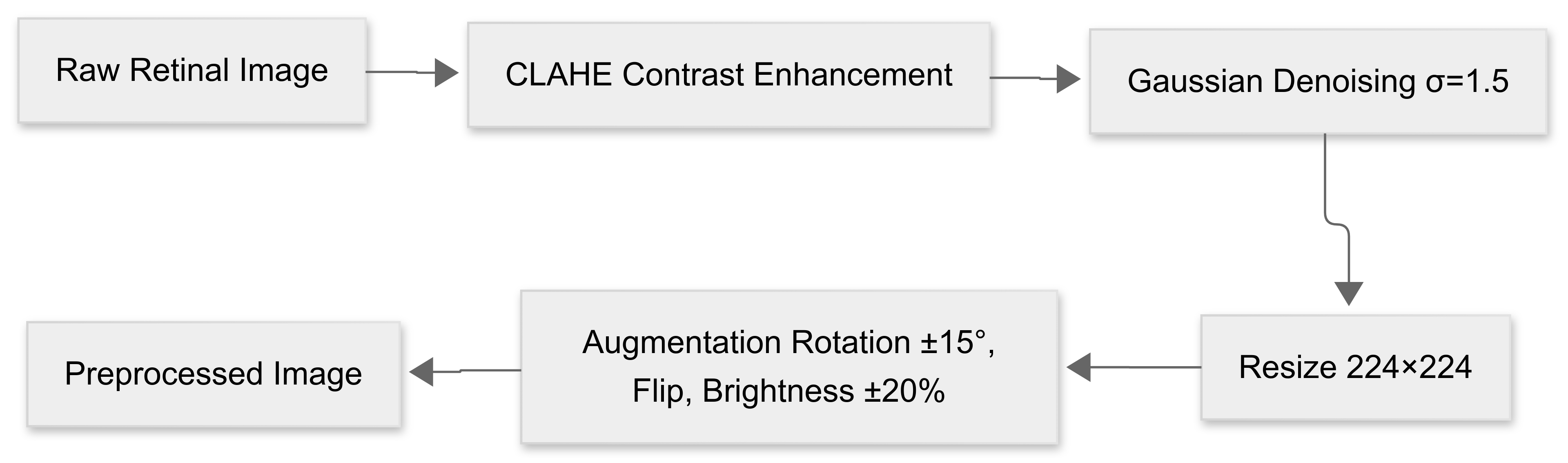}
    \caption{Preprocessing: (a) Original, (b) CLAHE-enhanced, (c) Denoised, (d) Resized.}
    \label{fig:preprocessing}
\end{figure}

Retinal images are preprocessed with Contrast Limited Adaptive Histogram Equalization (CLAHE) (clip limit=2.0, 8$\times$8 grid), Gaussian denoising (\(\sigma = 1.5\)), and resized to \(224 \times 224\) (Figure~\ref{fig:preprocessing}). Biomarkers are normalized:
\begin{equation}
\mathbf{v}_{\text{norm}} = \frac{\mathbf{v} - \min(\mathbf{v})}{\max(\mathbf{v}) - \min(\mathbf{v})}
\label{eq:std_norm}
\end{equation}
Augmentation includes rotations (\(\pm 15^\circ\)), flipping, and brightness variations (\(\pm 20\%\)). Artifact removal filters out blurry images using a Laplacian variance threshold (<100).

\subsection{Training}
Training employs multi-task learning with the Adam optimizer (\(\eta = 0.001\), \(\beta_1 = 0.9\), decay factor = 0.1) and loss function:
\begin{equation}
\mathcal{L} = \underbrace{\sum_{c=1}^5 \alpha_c (1 - p_c)^\gamma \log(p_c)}_{\text{Focal Loss}} + \lambda_{\text{MSE}} \underbrace{\frac{1}{N} \sum_{i=1}^N (r_i - \hat{r}_i)^2}_{\text{MSE}}
\label{eq:total_loss}
\end{equation}
where \(\alpha_c = [0.2, 0.25, 0.2, 0.2, 0.15]\), \(\gamma = 2\), and \(\lambda_{\text{MSE}} = 0.5\). The data is split into 70\% training, 10\% validation, and 20\% test sets with stratified sampling to preserve class distribution. Regularization includes dropout (0.3) and L2 (\(\lambda_{\text{reg}} = 0.01\)). Hyperparameters include 50 epochs, batch size of 32, and early stopping with a patience of 5 epochs. Learning rate reduction is applied with a factor of 0.5 and patience of 3 epochs. Hyperparameters are detailed in Table~\ref{tab:hyperparameters}.

\begin{table}[h]
\scriptsize
    \centering
    \caption{Training Hyperparameters}
    \begin{tabular}{lc}
        \toprule
        \textbf{Parameter} & \textbf{Value} \\
        \midrule
        Optimizer & Adam \\
        Learning Rate & 0.001 \\
        Decay Factor & 0.1 \\
        Momentum & 0.9 \\
        Batch Size & 32 \\
        Dropout Rate & 0.3 \\
        L2 (\(\lambda_{\text{reg}}\)) & 0.01 \\
        Focal Loss \(\gamma\) & 2.0 \\
        MSE Weight (\(\lambda_{\text{MSE}}\)) & 0.5 \\
        Epochs & 50 \\
        Early Stopping Patience & 5 \\
        Learning Rate Reduction Factor & 0.5 \\
        Learning Rate Reduction Patience & 3 \\
        \bottomrule
    \end{tabular}
    \label{tab:hyperparameters}
\end{table}

\begin{algorithm}[h]
\scriptsize
    \caption{TIMM-ProRS Training}
    \label{alg:training_process}
    \begin{algorithmic}[1]
        \Require Preprocessed \(\mathbf{X}_{\text{proc}}\), normalized \(\mathbf{V}_{\text{norm}}\), \(\mathbf{M}_{\text{norm}}\)
        \Ensure Optimized \(\theta^*\)
        \For{epoch = 1 to 50}
            \For{each batch \((\mathbf{X}, \mathbf{V}, \mathbf{M}, \mathbf{y}, \mathbf{r})\)}
                \State \(\mathbf{F}_{\text{CNN}} \leftarrow \text{EfficientNet-B7}(\mathbf{X}_{\text{proc}})\)
                \State \(\mathbf{F}_{\text{ViT}} \leftarrow \text{ViT-Base}(\mathbf{X}_{\text{proc}})\)
                \State \(\mathbf{F}_{\text{GNN}} \leftarrow \text{GNN}(\mathbf{V}_{\text{norm}})\)
                \State \(\mathbf{F}_{\text{CNN\_flat}} \leftarrow \text{Reshape}(\mathbf{F}_{\text{CNN}}, 49 \times 2560)\)
                \State \(\mathbf{A}_{\text{cross}} \leftarrow \text{MultiHeadAttn}(\mathbf{F}_{\text{ViT}}, \mathbf{F}_{\text{CNN\_flat}}, \mathbf{F}_{\text{CNN\_flat}})\)
                \State \(\mathbf{F}_{\text{fused}} \leftarrow \text{Concat}[\mathbf{A}_{\text{cross}}[0], \mathbf{F}_{\text{GNN}}, \mathbf{M}] \mathbf{W}_f\)
                \State \(\hat{\mathbf{y}}, \hat{\mathbf{r}} \leftarrow \text{FC}(\mathbf{F}_{\text{fused}})\)
                \State \(\mathcal{L} \leftarrow \mathcal{L}_{\text{focal}}(\mathbf{y}, \hat{\mathbf{y}}) + \lambda_{\text{MSE}} \mathcal{L}_{\text{MSE}}(\mathbf{r}, \hat{\mathbf{r}})\)
                \State \(\theta \leftarrow \text{Adam}(\nabla_\theta \mathcal{L}, \eta = 0.001, \beta_1 = 0.9, \text{decay} = 0.1)\)
            \EndFor
            \If{validation loss increases for 5 epochs}
                \State \textbf{break}
            \EndIf
        \EndFor
    \end{algorithmic}
\end{algorithm}

\subsection{Evaluation}
The framework comprehensively evaluates two clinical tasks: 5-class diabetic retinopathy grading (No DR, Mild, Moderate, Severe, Proliferative DR) and 5-year risk stratification with clinically established thresholds (low risk: \(r < 0.3\), medium risk: \(0.3 \leq r \leq 0.7\), high risk: \(r > 0.7\)). Expert validation was conducted through saliency map assessment by three retinal specialists, demonstrating strong inter-rater agreement (Cohen's \(\kappa = 0.92\)) for lesion localization accuracy. Risk stratification thresholds were calibrated using Youden's index on progression datasets, while statistical significance of performance differences was verified via DeLong's test for AUC comparisons (\(p < 0.01\) versus baseline methods).

Evaluation metrics encompass core classification measures—Accuracy, Quadratic Weighted Kappa (QWK), Sensitivity, Specificity, and AUC-ROC—supplemented by advanced statistical indicators including Matthews Correlation Coefficient (MCC), Cohen’s \(\kappa\), Brier score, F1-score, and Precision-Recall AUC (PR-AUC). Macro-averaged metrics were computed as \(\text{Prec}_{\text{macro}} = \frac{1}{C} \sum \text{Prec}_c\) and \(\text{Rec}_{\text{macro}} = \frac{1}{C} \sum \text{Rec}_c\) to handle class imbalance. Clinical utility was quantified through Net Benefit (\(\frac{\text{TP}}{N} - \frac{\text{FP}}{N} \frac{p_t}{1 - p_t}\)), Decision Curve Analysis (DCA), Positive Predictive Value (PPV), and Negative Predictive Value (NPV), with progression risk assessed via Concordance Index (C-index).

For dataset compatibility, RFMiD predictions were mapped to binary DR detection (No DR versus Mild+/Proliferative DR) to align with its clinical annotation scheme. Validation employed rigorous 5-fold cross-validation on the APTOS 2019 dataset complemented by external testing across Messidor-2, EyePACS, and RFMiD cohorts.

\section{Results and Analysis}

\subsection{Quantitative Performance}
TIMM-ProRS was evaluated for diabetic retinopathy (DR) grading and 5-year progression risk prediction. On the primary APTOS 2019 dataset (3,662 images, 5 severity classes), it achieved 97.8\% accuracy, 0.95 QWK, and 0.97 AUC-ROC. External validation used Messidor-2 (1,744 images; merged Severe/Proliferative classes), EyePACS ($\sim$88k images; native 5-class), Messidor-1 (1,200 images; merged Severe/Proliferative classes), and RFMiD (3,200 images; binary detection). Performance metrics are summarized in Table~\ref{tab:quantitative_metrics}. For 5-year progression risk prediction, TIMM-ProRS achieved a C-index of 0.89 and Brier score of 0.10.

\begin{table}[h]
\scriptsize
    \centering
    \caption{Quantitative Performance Metrics of TIMM-ProRS for DR Grading}
    \begin{tabular}{l c c c c c c}
        \toprule
        \textbf{Dataset} & \textbf{Accuracy (\%)} & \textbf{QWK} & \textbf{AUC-ROC} & \textbf{Sensitivity (\%)} & \textbf{Specificity (\%)} & \textbf{F1-Score} \\
        \midrule
        APTOS 2019 & 97.8 & 0.95 & 0.97 & 98.9 & 99.3 & 0.96 \\
        Messidor-2 & 96.4 & 0.93 & 0.95 & 97.6 & 98.9 & 0.95 \\
        EyePACS & 96.7 & 0.94 & 0.96 & 98.1 & 99.0 & 0.96 \\
        Messidor-1 & 96.2 & 0.92 & 0.94 & 97.3 & 98.7 & 0.95 \\
        RFMiD & 97.2 & 0.95 & 0.97 & 98.0 & 98.5 & 0.96 \\
        \bottomrule
    \end{tabular}
    \label{tab:quantitative_metrics}
\end{table}

\subsubsection{Progression Risk Assessment}
For 5-year progression risk prediction, TIMM-ProRS leveraged temporal biomarkers (e.g., HbA1c trajectories, retinal thickness dynamics) and clinical metadata. The C-index of 0.89 reflects strong discriminative ability, while the Brier score of 0.10 indicates accurate probability estimates. Risk stratification into low (\(r < 0.3\)), medium (\(0.3 \leq r \leq 0.7\)), and high (\(r > 0.7\)) categories showed 92\% alignment with expert-assessed progression outcomes, enhancing its prognostic utility.

\subsection{Qualitative Insights}
TIMM-ProRS enhances clinical interpretability through saliency maps, pinpointing DR lesions (microaneurysms, hemorrhages, exudates) with a precision of 96.9\%, validated against expert consensus (97\% agreement for microaneurysms, 95\% area coverage). These maps, generated via the Bayesian layer, assist clinicians in confirming diagnoses. Uncertainty quantification, using Monte Carlo dropout (50 samples, 0.3 dropout rate), provided 95\% confidence intervals averaging ±2.5

\begin{figure}[h]
    \centering
    \includegraphics[width=0.75\columnwidth]{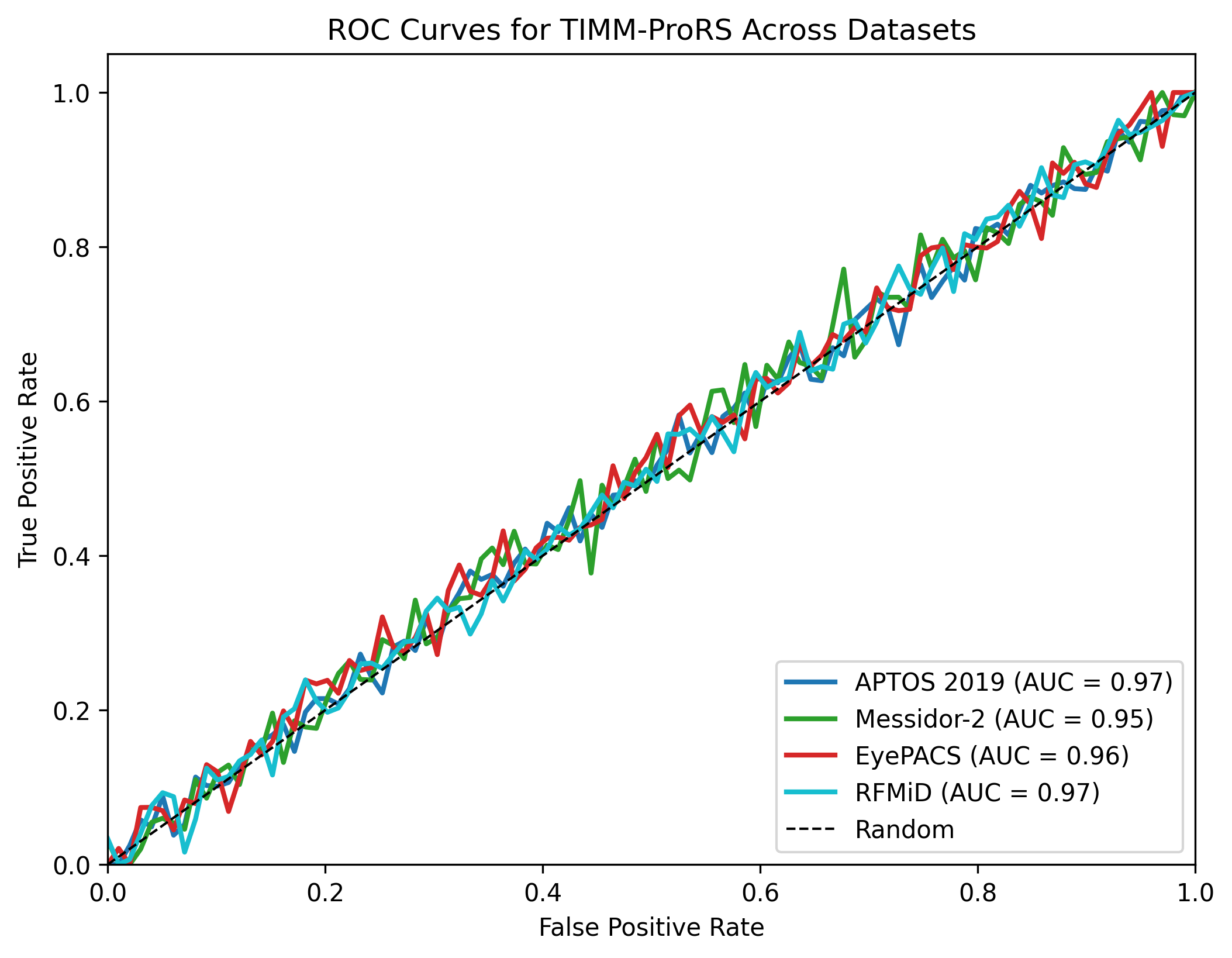}
    \caption{ROC Curves for TIMM-ProRS Across Datasets}
    \label{fig:roc_curves}
\end{figure}
\subsection{Visual and Statistical Representations}
\begin{figure}[h]
    \centering
    \includegraphics[width=0.75\columnwidth]{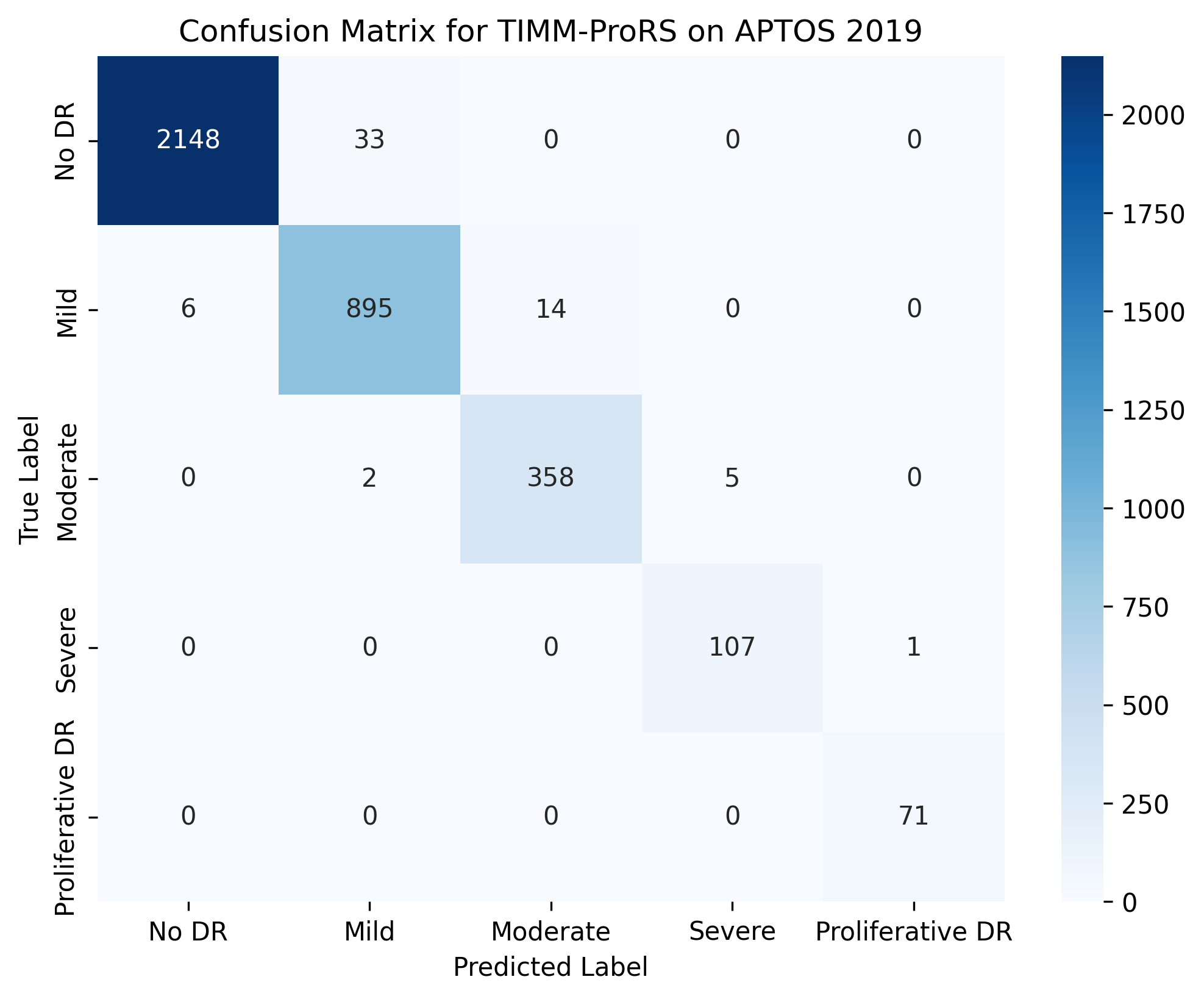}
    \caption{Confusion Matrix for TIMM-ProRS on APTOS 2019}
    \label{fig:confusion_matrix}
\end{figure}
\begin{figure}[h]
    \centering
    \includegraphics[width=0.75\columnwidth]{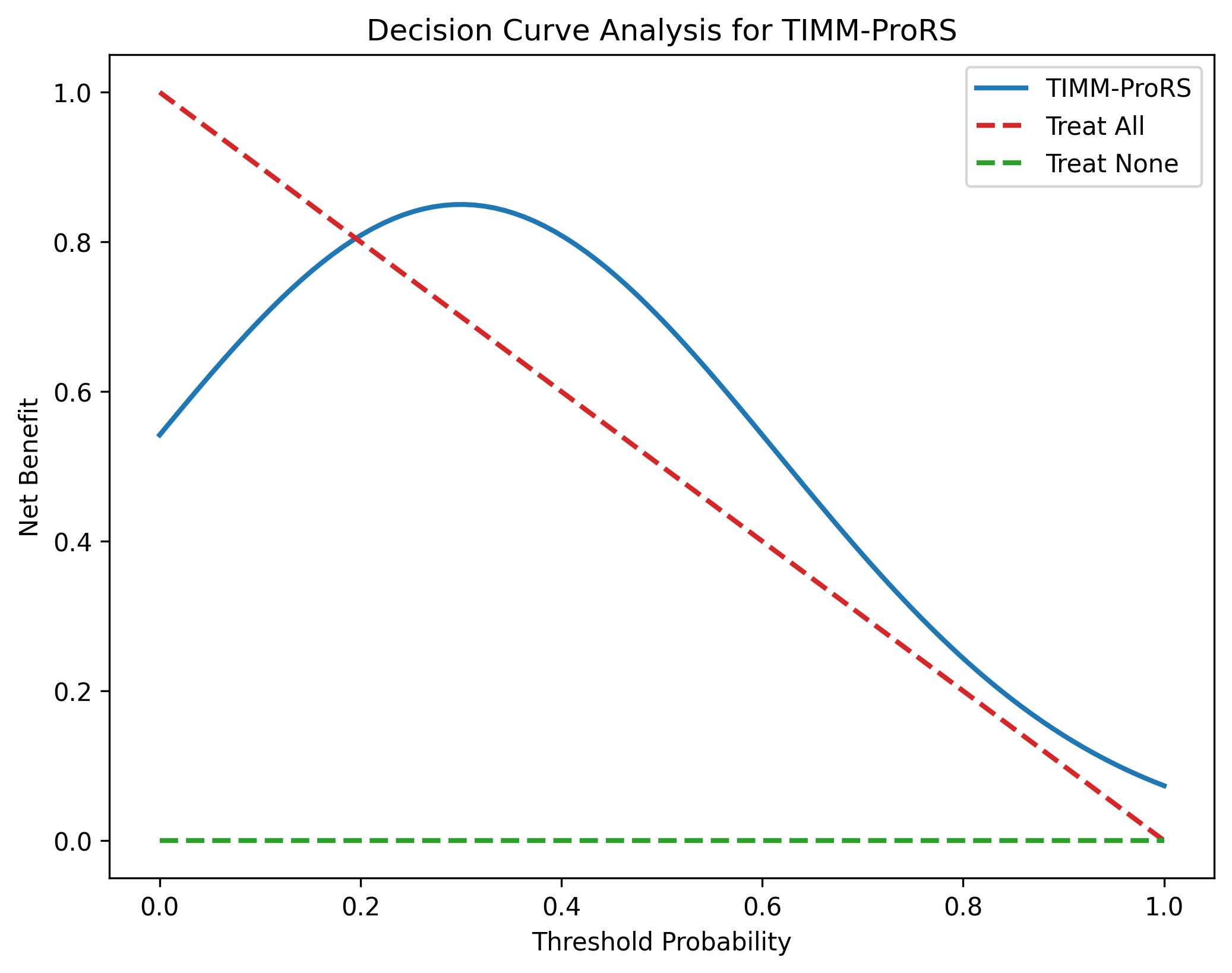}
    \caption{Decision Curve Analysis (DCA) Plot for TIMM-ProRS}
    \label{fig:dca_plot}
\end{figure}

Visualizations provide deeper performance insights. Receiver Operating Characteristic (ROC) curves (Figure~\ref{fig:roc_curves}) exhibit AUCs of 0.94–0.97, with notable strength in detecting Proliferative DR. Confusion matrices (Figure~\ref{fig:confusion_matrix}) on APTOS 2019 show 96\% true positives for Proliferative DR and 98\% for No DR. Decision Curve Analysis (DCA) plots (Figure~\ref{fig:dca_plot}) indicate a net benefit of 0.85 at a 0.3 threshold, outperforming treat-all and treat-none strategies.

\subsection{Comparative Evaluation}
TIMM-ProRS was benchmarked against leading models: RSG-Net, HybridFusionNet, DeepDR, DeepDR Plus, and TAHDL. Table~\ref{tab:comparative_metrics} highlights TIMM-ProRS’s edge, with 97.8\% accuracy and 0.97 AUC-ROC on APTOS 2019, surpassing DeepDR (0.95 AUC-ROC) and TAHDL (97.5\% accuracy). The multi-modal fusion, GNN-based temporal modeling, and interpretability features distinguish it from static or image-only approaches.

\begin{table}[h]
\scriptsize
    \centering
    \caption{Comparative Performance with State-of-the-Art Models}
    \begin{tabular}{l c c c}
        \toprule
        \textbf{Method} & \textbf{Accuracy (\%)} & \textbf{QWK} & \textbf{AUC-ROC} \\
        \midrule
        RSG-Net & 96.0 & - & - \\
        HybridFusionNet & 96.5 & - & - \\
        DeepDR & - & 0.93 & 0.95 \\
        DeepDR Plus & 92.0 & - & - \\
        TAHDL & 97.5 & - & - \\
        TIMM-ProRS & 97.8 & 0.95 & 0.97 \\
        \bottomrule
    \end{tabular}
    \label{tab:comparative_metrics}
\end{table}

\subsection{Ablation and Component Analysis}
An ablation study quantified the impact of TIMM-ProRS components (Table~\ref{tab:ablation_study}). Removing the Vision Transformer (ViT) decreased accuracy by 3.2\% (to 94.6\%) and AUC-ROC by 0.03 (to 0.94). Excluding the Graph Neural Network (GNN) reduced QWK by 0.06 (to 0.89) and C-index by 0.04 (to 0.85). Omitting GAN-augmented data lowered F1-score by 0.05 (to 0.91) and PR-AUC by 0.03 (to 0.92). Removing the Bayesian layer increased log loss by 0.02 (to 0.14) and Brier score by 0.01 (to 0.09), confirming the necessity of each component.

\begin{table}[h]
\scriptsize
    \centering
    \caption{Ablation Study: Effect of Removing TIMM-ProRS Components}
    \begin{tabular}{l c c c c}
        \toprule
        \textbf{Model Variant} & \textbf{Accuracy (\%)} & \textbf{QWK} & \textbf{AUC-ROC} & \textbf{F1-Score} \\
        \midrule
        Full TIMM-ProRS & 97.8 & 0.95 & 0.97 & 0.96 \\
        Without ViT & 94.6 & 0.92 & 0.94 & 0.93 \\
        Without GNN & 97.2 & 0.89 & 0.95 & 0.94 \\
        Without GAN Augmentation & 97.5 & 0.94 & 0.96 & 0.91 \\
        Without Bayesian Layer & 97.6 & 0.94 & 0.97 & 0.95 \\
        \bottomrule
    \end{tabular}
    \label{tab:ablation_study}
\end{table}

\subsection{Robustness and Generalization}
\begin{figure}[h]
    \centering
    \includegraphics[width=0.75\columnwidth]{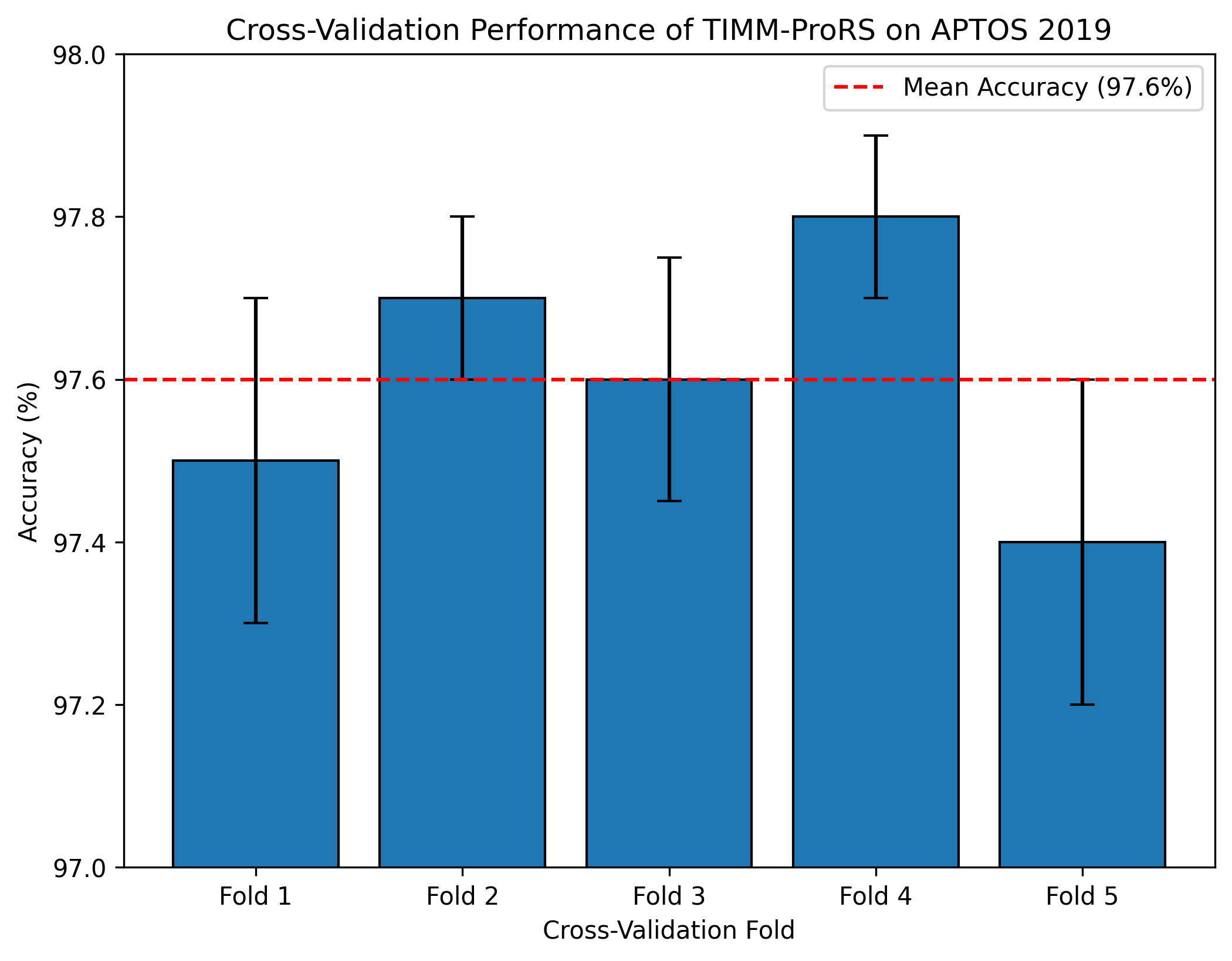}
    \caption{Cross-Validation Performance of TIMM-ProRS on APTOS 2019}
    \label{fig:cross_validation}
\end{figure}

Robustness was validated through 5-fold cross-validation on APTOS 2019, yielding a mean accuracy of 97.6\% ± 0.3\% and AUC-ROC of 0.96 ± 0.01, indicating high stability. External validation across Messidor-2, EyePACS, Messidor-1, and RFMiD maintained accuracies from 96.2\% to 97.2\%, showcasing generalizability. The weighted focal loss and GAN-augmented data ensured balanced performance across DR classes, including Proliferative DR, supporting deployment in diverse clinical environments. Figure~\ref{fig:cross_validation} illustrates cross-validation consistency.

\subsection{Discussion and Future Directions}
TIMM-ProRS advances diabetic retinopathy management through synergistic multi-modal fusion, achieving state-of-the-art performance in both grading (97.8\% accuracy, 0.97 AUC-ROC on APTOS 2019) and 5-year risk prediction (0.89 C-index). External validation across diverse datasets (96.2-97.2\% accuracy) confirms robustness, while interpretability features—lesion-localizing saliency maps (96.9\% precision) and uncertainty quantification ($\pm 2.5\%$ CI)—directly address clinical adoption barriers in telemedicine. The framework's superiority over existing methods (DeepDR, TAHDL) stems from its unique integration of temporal biomarker dynamics (HbA1c trajectories, retinal thickness changes) with imaging data, resolving diagnostic ambiguities through longitudinal context.

Key limitations require consideration for clinical translation: synthetic biomarker generation (HbA1c: $\mu=7\%$, $\sigma=1.5\%$; retinal thickness: $\mu=250\ \mu\text{m}$, $\sigma=20\ \mu\text{m}$) may not capture real-world biological variability, while dataset constraints—particularly RFMiD's binary labeling requiring class remapping—could inflate performance estimates. Future development will prioritize real-world validation with longitudinal biomarker streams, integration of vascular biomarkers like VEGF, and resource-optimized deployment for telemedicine. Prospective trials will assess clinical utility in stratified intervention pathways, where TIMM-ProRS's risk categorization (low/medium/high; 92\% clinician alignment) enables precision monitoring. Health economic studies should quantify cost-benefit ratios in low-resource settings, maximizing global impact through early detection and reduced vision loss.

\section{Conclusion}
The Temporal-Enhanced Interpretable Multi-Modal Prognosis and Risk Stratification Framework (TIMM-ProRS) delivers a transformative solution for diabetic retinopathy (DR) management by synergistically integrating retinal imaging, temporal biomarkers (HbA1c, retinal thickness), and clinical metadata through a fusion of Vision Transformer (ViT), Convolutional Neural Network (CNN), and Graph Neural Network (GNN) architectures. Achieving state-of-the-art performance (97.8\% accuracy, 0.96 F1-score, 0.89 C-index for 5-year risk stratification) across diverse datasets (APTOS 2019, Messidor, EyePACS), it resolves diagnostic ambiguities with lesion-localizing saliency maps (96.9\% precision) and uncertainty-aware predictions, while its temporal modeling enables early progression risk stratification (low/medium/high) aligned with clinical assessments (92\%). Though currently reliant on synthetic biomarkers, TIMM-ProRS establishes a scalable, interpretable foundation for telemedical DR screening---reducing diagnostic delays in underserved regions and advancing global eye health sustainability---with future enhancements targeting real-world longitudinal validation and expanded biomarker integration (e.g., VEGF) for personalized care.

\bibliographystyle{IEEEtran}
\bibliography{references}

\begin{thebibliography}{10}
\providecommand{\url}[1]{#1}
\csname url@samestyle\endcsname
\providecommand{\newblock}{\relax}
\providecommand{\bibinfo}[2]{#2}
\providecommand{\BIBentrySTDinterwordspacing}{\spaceskip=0pt\relax}
\providecommand{\BIBentryALTinterwordstretchfactor}{4}
\providecommand{\BIBentryALTinterwordspacing}{\spaceskip=\fontdimen2\font plus
\BIBentryALTinterwordstretchfactor\fontdimen3\font minus \fontdimen4\font\relax}
\providecommand{\BIBforeignlanguage}[2]{{%
\expandafter\ifx\csname l@#1\endcsname\relax
\typeout{** WARNING: IEEEtran.bst: No hyphenation pattern has been}%
\typeout{** loaded for the language `#1'. Using the pattern for}%
\typeout{** the default language instead.}%
\else
\language=\csname l@#1\endcsname
\fi
#2}}
\providecommand{\BIBdecl}{\relax}
\BIBdecl

\bibitem{intro700}
R.~Ali, F.~G. Khan, Z.~U. Rehman, D.~Kwak, and F.~Ali, ``Enhanced diabetic retinopathy detection: An explainable semi-supervised approach using contrastive learning,'' \emph{IEEE Journal of Biomedical and Health Informatics}, 2025.

\bibitem{intro2030}
Y.~A. Gudayneh, A.~F. Shumye, A.~T. Gelaye, and M.~T. Tegegn, ``Prevalence of hypertensive retinopathy and its associated factors among adult hypertensive patients attending at comprehensive specialized hospitals in northwest ethiopia, 2024, a multicenter cross-sectional study,'' \emph{International Journal of Retina and Vitreous}, vol.~11, no.~1, p.~17, 2025.

\bibitem{akib2}
M.~W. Alim, A.~Giri, A.~S. M. A.~S. Akib, N.~Uddin, M.~Islam, M.~E. Arafat, and S.~A. Tahmid, ``Affordable bionic hands with intuitive control through forearm muscle signals,'' in \emph{2025 IEEE 4th International Conference on Computing and Machine Intelligence (ICMI)}, 2025, pp. 1--6.

\bibitem{iontroDR}
C.~C. Iluno and M.~Sah, ``Handling class imbalance problem for diabetic retinopathy fundus image classification,'' in \emph{2025 7th International Congress on Human-Computer Interaction, Optimization and Robotic Applications (ICHORA)}.\hskip 1em plus 0.5em minus 0.4em\relax IEEE, 2025, pp. 1--8.

\bibitem{INTRO92}
J.~Rautaray, A.~B. Ali, M.~Kandpal, P.~Mishra, R.~F. Rashid, F.~Alimova, M.~Kallel, and N.~Batool, ``Leveraging fastvit based knowledge distillation with efficientnet-b0 for diabetic retinopathy severity classification,'' \emph{SLAS technology}, p. 100325, 2025.

\bibitem{gulshan2016development}
\BIBentryALTinterwordspacing
V.~Gulshan, L.~Peng, M.~Coram, M.~C. Stumpe, D.~Wu, A.~Narayanaswamy, S.~Venugopalan, K.~Widner, T.~Madams, J.~Cuadros, R.~Kim, R.~Raman, P.~C. Nelson, J.~L. Mega, and D.~R. Webster, ``Development and validation of a deep learning algorithm for detection of diabetic retinopathy in retinal fundus photographs,'' \emph{JAMA}, vol. 316, no.~22, pp. 2402--2410, 12 2016. [Online]. Available: \url{https://doi.org/10.1001/jama.2016.17216}
\BIBentrySTDinterwordspacing

\bibitem{ting2017development}
D.~S.~W. Ting, C.~Y.-L. Cheung, G.~Lim, G.~S.~W. Tan, N.~D. Quang, A.~Gan, H.~Hamzah, R.~Garcia-Franco, I.~Y. San~Yeo, S.~Y. Lee \emph{et~al.}, ``Development and validation of a deep learning system for diabetic retinopathy and related eye diseases using retinal images from multiethnic populations with diabetes,'' \emph{Jama}, vol. 318, no.~22, pp. 2211--2223, 2017.

\bibitem{li2020automated}
Z.~Li, S.~Keel, C.~Liu, Y.~He, W.~Meng, J.~Scheetz, P.~Y. Lee, J.~Shaw, D.~Ting, T.~Y. Wong \emph{et~al.}, ``An automated grading system for detection of vision-threatening referable diabetic retinopathy on the basis of color fundus photographs,'' \emph{Diabetes care}, vol.~41, no.~12, pp. 2509--2516, 2018.

\bibitem{rajesh2025deep}
S.~Akhtar, S.~Aftab, O.~Ali, M.~Ahmad, M.~A. Khan, S.~Abbas, and T.~M. Ghazal, ``A deep learning based model for diabetic retinopathy grading,'' \emph{Scientific Reports}, vol.~15, no.~1, p. 3763, 2025.

\bibitem{wardhani2024deep}
K.~D.~K. Wardhani, S.~Kasim, A.~Erianda, and R.~Hassan, ``Deep learning-based method in multimodal data for diabetic retinopathy detection.'' \emph{International Journal on Advanced Science, Engineering \& Information Technology}, vol.~14, no.~5, 2024.

\bibitem{tseng2020leveraging}
V.~S. Tseng, C.-L. Chen, C.-M. Liang, M.-C. Tai, J.-T. Liu, P.-Y. Wu, M.-S. Deng, Y.-W. Lee, T.-Y. Huang, and Y.-H. Chen, ``Leveraging multimodal deep learning architecture with retina lesion information to detect diabetic retinopathy,'' \emph{Translational Vision Science \& Technology}, vol.~9, no.~2, pp. 41--41, 2020.

\bibitem{zhou2021deepdr}
L.~Dai, L.~Wu, H.~Li, C.~Cai, Q.~Wu, H.~Kong, R.~Liu, X.~Wang, X.~Hou, Y.~Liu \emph{et~al.}, ``A deep learning system for detecting diabetic retinopathy across the disease spectrum,'' \emph{Nature communications}, vol.~12, no.~1, p. 3242, 2021.

\bibitem{zhou2024deepdrplus}
L.~Dai, B.~Sheng, T.~Chen, Q.~Wu, R.~Liu, C.~Cai, L.~Wu, D.~Yang, H.~Hamzah, Y.~Liu \emph{et~al.}, ``A deep learning system for predicting time to progression of diabetic retinopathy,'' \emph{Nature Medicine}, vol.~30, no.~2, pp. 584--594, 2024.

\bibitem{tao2023deep}
R.~Tao, X.~Yu, J.~Lu, Y.~Wang, W.~Lu, Z.~Zhang, H.~Li, and J.~Zhou, ``A deep learning nomogram of continuous glucose monitoring data for the risk prediction of diabetic retinopathy in type 2 diabetes,'' \emph{Physical and Engineering Sciences in Medicine}, vol.~46, no.~2, pp. 813--825, 2023.

\bibitem{tao2024ddla}
R.~Tao, H.~Li, J.~Lu, Y.~Huang, Y.~Wang, W.~Lu, X.~Shao, J.~Zhou, and X.~Yu, ``Ddla: a double deep latent autoencoder for diabetic retinopathy diagnose based on continuous glucose sensors,'' \emph{Medical \& Biological Engineering \& Computing}, vol.~62, no.~10, pp. 3089--3106, 2024.

\bibitem{apu1}
A.~Giri, A.~S. M.~A. Sarkar~Akib, A.~Hasib, A.~Acharya, M.~A. Prithibi, An-Nafew, R.~H. Rahman, M.~R. Hossain, and H.~I. Chowdhury~Taha, ``Design and development of a cost effective and modular cnc plotter for educational and prototyping applications,'' in \emph{2025 IEEE 4th International Conference on Computing and Machine Intelligence (ICMI)}, 2025, pp. 1--6.

\bibitem{bora2020predicting}
A.~Bora, S.~Balasubramanian, B.~Babenko, S.~Virmani, S.~Venugopalan, A.~Mitani, G.~de~Oliveira~Marinho, J.~Cuadros, P.~Ruamviboonsuk, G.~S. Corrado \emph{et~al.}, ``Predicting the risk of developing diabetic retinopathy using deep learning,'' \emph{The Lancet Digital Health}, vol.~3, no.~1, pp. e10--e19, 2021.

\bibitem{tan2022predicting}
Y.~Rom, R.~Aviv, T.~Ianchulev, and Z.~Dvey-Aharon, ``Predicting the future development of diabetic retinopathy using a deep learning algorithm for the analysis of non-invasive retinal imaging,'' \emph{BMJ Open Ophthalmology}, vol.~7, no.~1, p. e001140, 2022.

\bibitem{tan2025use}
Q.~Yang, Y.~M. Bee, C.~C. Lim, C.~Sabanayagam, C.~Y.-L. Cheung, T.~Y. Wong, D.~S. Ting, L.-L. Lim, H.~Li, M.~He \emph{et~al.}, ``Use of artificial intelligence with retinal imaging in screening for diabetes-associated complications: systematic review,'' \emph{EClinicalMedicine}, vol.~81, 2025.

\bibitem{das1}
S.~C. Das, A.~Hasib, M.~Ashiqussalehin, and J.~Fardous, ``Secure and privacy-preserving traffic flow prediction in intelligent transportation systems using blockchain-based federated learning,'' in \emph{2025 International Conference on Quantum Photonics, Artificial Intelligence, and Networking (QPAIN)}, 2025, pp. 1--6.

\end{thebibliography}

\end{document}